\documentclass{svproc}
\usepackage{url}
\usepackage{graphicx}
\usepackage{marginnote}
\usepackage{algorithm}
\usepackage{algpseudocode}
\usepackage{amsmath}
\usepackage{dirtytalk}
\usepackage{xcolor}
\usepackage{subfig}

\usepackage[OT2,OT1]{fontenc}
\usepackage[russian,english]{babel}

\begin{document}
\mainmatter              
\title{Change my Mind: Data Driven Estimate of Open-Mindedness from Political Discussions}
\titlerunning{Data Driven Estimate of
Open-Mindedness from Political Discussions}  
%
\author{Valentina Pansanella\inst{1} \and Virginia Morini\inst{2} \and Tiziano Squartini\inst{3} \and Giulio Rossetti\inst{4}}
\authorrunning{Valentina Pansanella et al.} 
%
\tocauthor{Valentina Pansanella, Virginia Morini, Tiziano Squartini and Giulio Rossetti}
\institute{Scuola Normale Superiore, Pisa, Italy\\
\email{valentina.pansanella@sns.it}
\and
Computer Science department, University of Pisa, Italy\\
\email{virginia.morini@phd.unipi.it}
\and
IMT School for Advanced Studies, Lucca, Italy\\
\email{tiziano.squartini@imtlucca.it}
\and
KDD Laboratory, ISTI, National Research Council, Pisa, Italy\\
\email{giulio.rossetti@isti.cnr.it}}

\maketitle              

\begin{abstract}
One of the main dimensions characterizing the unfolding of opinion formation processes in social debates is the degree of open-mindedness of the involved population.
Opinion dynamic modeling studies have tried to capture such a peculiar expression of individuals' personalities and relate it to emerging phenomena like polarization, radicalization, and ideology fragmentation.
However, one of their major limitations lies in the strong assumptions they make on the initial distribution of such characteristics, often fixed so as to satisfy a normality hypothesis.

Here we propose a data-driven methodology to estimate users' open-mindedness from online discussion data.
Our analysis - focused on the political discussion taking place on Reddit during the first two years of the Trump presidency - unveils the existence of statistically diverse distributions of open-mindedness in annotated sub-populations (i.e., Republicans, Democrats, and Moderates/Neutrals).
Moreover, such distributions appear to be stable across time and generated by individual users' behaviors that remain consistent and underdispersed.

\keywords{Online Debates, Open-Mindedness, Political Polarization, Opinion Dynamics}
\end{abstract}

\section{Introduction}
One of the most debated and analyzed phenomena on online social networks is the tendency to observe political polarization \cite{conover2011political,morales2015measuring,bakshy2015exposure,morini2020capturing}, i.e., the divergence of political attitudes to ideological extremes not aiming at reaching any form of synthesis.
Indeed, social media blurred the boundaries of communication and democratized content diffusion. However, the fact that we can potentially engage with a lot of different information does not mean that we can engage with and be influenced by opposing or even mildly different stances \cite{pariser2011filter}. 

Furthermore, humans are far from being perfectly rational individuals and are affected by a series of cognitive biases\footnote[1]{Cognitive biases cheat sheet: \url{shorturl.at/afjpM}, last visited June 2022.} in the process of forming their belief system. 
Among these, it is worth mentioning confirmation bias \cite{nickerson1998confirmation,kunda1990case},  the human tendency to ignore content that counters their prior beliefs by a) choosing to interact with like-minded individuals \cite{MSL01} b) ignoring interactions with the ``opposite faction" \cite{Deffuant2000MixingBA}. 
Unfortunately, by connecting with like-minded individuals and ignoring contrasting views, users can easily compromise their open-mindedness \cite{verducci2019critical}, exacerbating and polarizing their views even more, and, in the end, giving rise to the formation and persistence of echo chambers \cite{10.5555/1536923}. 

One of the most exploited approaches for understanding the effects of different kinds of biases on public opinions - especially political opinions - is through mathematical models of opinion formation \cite{sirbu2017opinion}, where parameters incorporate psychological factors (e.g., cognitive biases) affecting individual opinion evolution. 
Despite having some advantages, opinion dynamics models lack empirical validation \cite{surveyjanos}.
However, thanks to the advent of the Internet - and with the rise of social media - an increasing part of human interactions leave a massive digital footprint that can be exploited to study the dynamics of opinion formation and diffusion. Following such reasoning, since different models or different parameter values can predict different, even opposite, effects of biases on opinions \cite{Maes2015WillTP}, there is a crucial need for empirical works to study and quantify socio-psychological and external drivers of the dynamics. 

In this work, we move in this direction by studying how being politically polarized affects users' open-mindedness over time. 
In detail, we focus on a twenty-months discussion on {\fontfamily{qcr}\selectfont r/politics} (i.e., the largest political subreddit on Reddit), computing the probability of changing one's political leaning over time and estimating the distribution of open-mindedness within the user population and its temporal evolution. 
Firstly, this analysis aims to understand how open-mindedness is distributed within the population and over time. A second aim is to understand if there is a substantial difference in the level of open-mindedness in relation to the political leaning, i.e., if more extreme individuals are more close-minded than moderate ones or vice versa. 
Our results suggest that the different subpopulations' open-mindedness distributions are stable in time and statistically different: Moderates/Neutrals and Republicans showing more nuanced close-mindedness patterns w.r.t. Democrats.
Finally, our analysis also underlines how Reddit users tend to be consistent in their open-mindedness attitude across time, showing, on average, low degrees of variance and dispersion.

The rest of the paper is organized as follows. 
In Section \ref{sec:related}, we discuss the literature
involving open-mindedness modeling both from an opinion dynamics and data-driven point of view. 
Section \ref{sec:data} discusses how we leverage Reddit data in order to model opinion evolution over time. 
Then, in Section \ref{sec:stab_and_od}, we propose an algorithm to estimate users' open-mindedness and we longitudinally study its evolution in three well-characterized sub-populations. 
Finally, Section \ref{sec:conclusions} concludes the paper with a brief summary of the main results and discusses limitations and directions for future work.

\section{Related work}
\label{sec:related}
In the process of opinion formation cognitive biases play a role because people usually feel discomfort when encountering divergent opinions, thus tending to select information that confirm their pre-existing beliefs \cite{nickerson1998confirmation}. 
One of the first attempts at understanding the effects of such kind of cognitive bias is the model by Deffuant, and Weisbuch \cite{Deffuant2000MixingBA}. 
In this case, the bias is introduced with a single parameter, namely ``confidence bound”, which indicates that agents only influence each other if their opinion distance is below a certain threshold. 
The main result of this model is that if the population is ``close-minded", polarization arises. 
So it appears that cognitive biases may be enough to justify opinion polarization, even without considering additional factors, such as the network structure or the information environment. 
However, the effects of open-mindedness alone and in interplay with other factors have been extensively studied through opinion dynamics models, extending the baseline \cite{Deffuant2000MixingBA} with additional agent-level characteristics \cite{lorenz2010heterogeneous}, considering multi-dimensional opinions \cite{Deffuant2013TheLM} or adding repulsive behaviors \cite{Chen2021OpinionDW}.
Despite conclusions from these models appear to be realistic and seem to explain why polarization arises and resists in many social networks, there is a need for empirical works unveiling the level of open-mindedness within polarizing topics.
Indeed, to the best of our knowledge, there are still no works leveraging the vast amount of big data traces left from online social networks to tackle such a task. 
With respect to modelling works, simulation results have been compared to election data \cite{Vendeville2021ForecastingER}, survey data \cite{Sobkowicz2016QuantitativeAB}, social experiments data \cite{Stewart2019InformationGA}, or social media data \cite{Xiong2014OpinionFO} in order to manually tune parameters or learn them with automated approaches \cite{monti2020learning,sichani2017inference,kulkarni2017slant+,papercinelli}. 
Different data-driven works \cite{gilbert2009blogs,hilbert2018communicating,ge2020understanding,nguyen2014exploring,quattrociocchi2016echo,bakshy2015exposure} have shown how the polarization of opinions around political issues led the people to enter in closed polarized systems insulated from rebuttal, i.e., echo chambers.
Interestingly, authors in \cite{garimella2018political} found that echo chambers do not arise around topics not politically polarized (e.g., TV shows, food), thus suggesting that the echo chamber phenomenon and users' open-mindedness might be strictly related to the discussed topic.






\section{Analyzing the Political Debate Online}\label{sec:data}
\begin{figure}[t]
     \centering
     \subfloat[]{\includegraphics[width=0.5\textwidth]{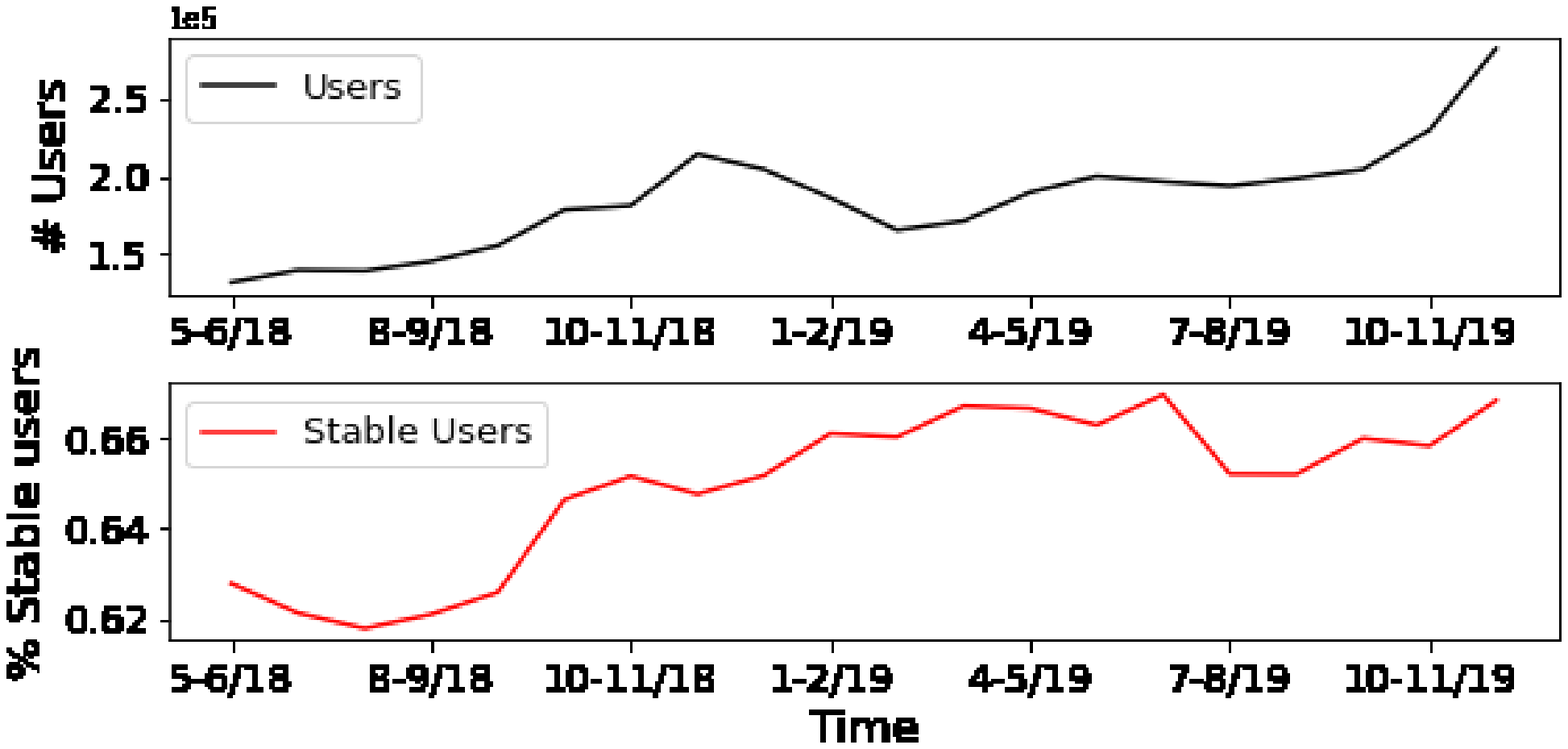}\label{img:multi2}}
     \subfloat[]{\includegraphics[width=0.45\textwidth]{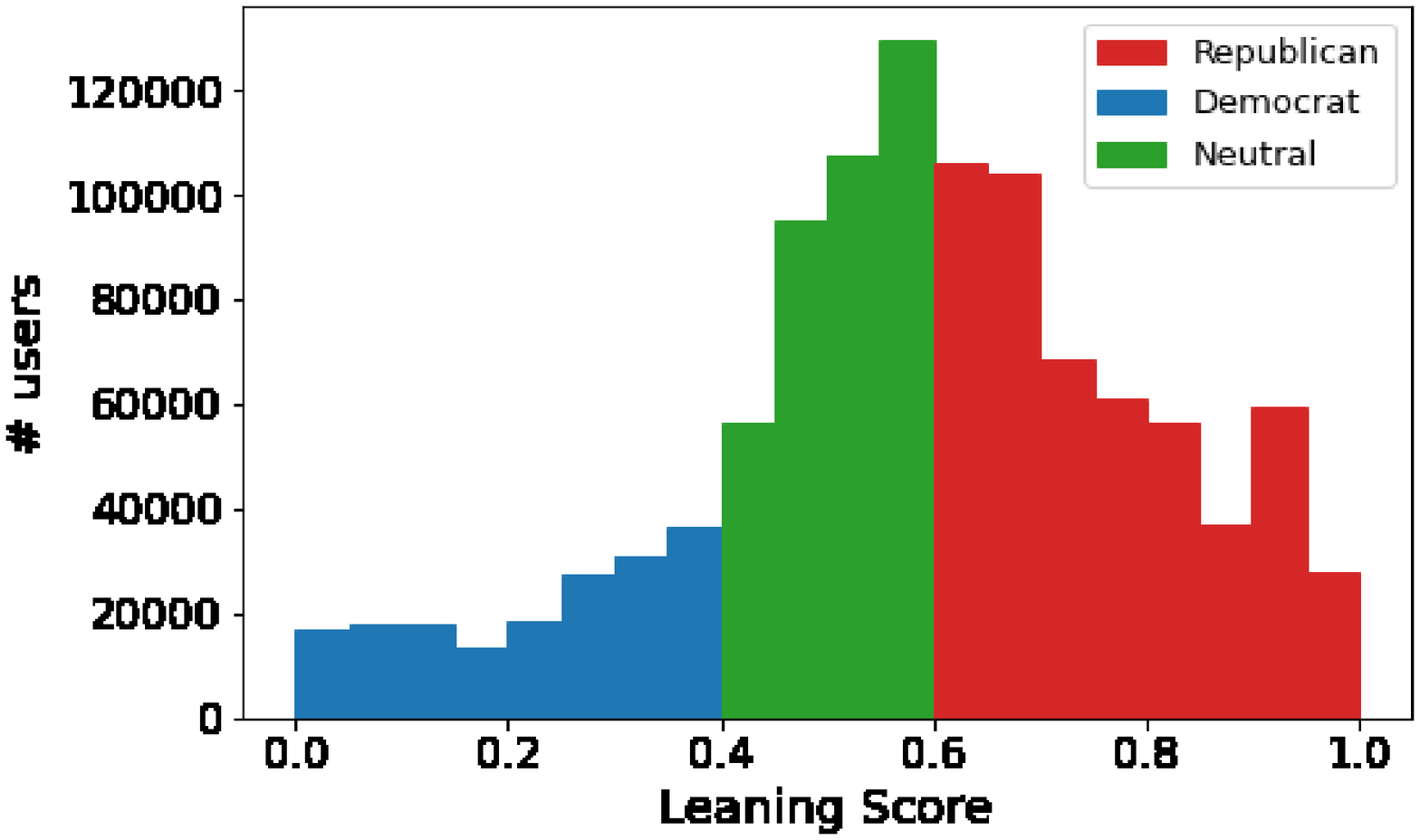}}
     \caption{a) Top: For each month, the number of users who participated in the debate. Bottom: For each month, the percentage of users that are stable across contiguous months. b) Authors' leaning distribution in the whole time period.}
     \label{fig:user_leaning}
 \end{figure}

\noindent{\bf Data Collection.}
For the purpose of this work, we built a dataset of political discussions on Reddit. 
Reddit is a popular social platform that allows users to post content to individual forums called subreddits, each dedicated to a specific topic. 
Such a categorized structure makes it easy to find users involved in specific debates. 
To assess the open-mindedness of users over time, we decide to select a quite controversial domain, i.e., Politics.   
Among the thousands of subreddits talking about politics we choose {\fontfamily{qcr}\selectfont r/politics}\footnote[2]{\label{note1}\url{https://www.reddit.com/r/politics/} ``{\fontfamily{qcr}\selectfont r/politics} is for news and discussion about US politics.''}, since it is the largest political subreddit and further, it is not aligned with a specific ideology but rather is visited regularly by users having different political beliefs. 
Notice that, as highlighted in the subreddit description\footnotemark[1], {\fontfamily{qcr}\selectfont r/politics} mainly refers to political discussion in the US. 
Thanks to the Pushshift API \cite{baumgartner2020pushshift}, we collected all posts and comments shared on the subreddit from May 2018 to December 2019, i.e., about one year and a half of Donald Trump's presidency - covering all discussions of 1,089795 users.
As shown in figure \ref{fig:user_leaning}a, the numbers of users who participated in the debate tend to increase over time, and further, more than 60\% of users are stable across contiguous months, meaning that they continue posting or commenting at least for two consecutive months.
The code and the data used for this work are available in a dedicated GitHub repository\footnote[3]{\url{https://github.com/ValentinaPansanella/OpenMindednessReddit.git}}.
\\ \ \\
\noindent{\bf Ideology Estimate.}
To assess if the open-mindedness of users evolves over time, we have to establish the ideology of users in different time periods. 
Since we are dealing with users debating political issues in the US, we try to categorize them with respect to the US two-party system: \textit{Republican} and \textit{Democratic}. \newline
For such a purpose, we model the task of predicting users' political alignment as a text classification problem. 
In other words, we leverage users' posts to measure their degree of alignment with Republican and Democrat ideologies.
To accomplish this task, we leverage an LSTM model that we trained on Reddit US political texts in our previous works \cite{morini2020,morini2021toward}. 
In detail, to train such a model we defined a ground truth composed of Reddit posts belonging to subreddits known to be either Pro-Trump or Anti-Trump (i.e., {\fontfamily{qcr}\selectfont r/The\_Donald} for the first group and {\fontfamily{qcr}\selectfont r/Fuckthealtright} and {\fontfamily{qcr}\selectfont r/EnoughTrumpSpam} for the second). 
Accordingly, we modeled the text classification task as a binary problem.  During model selection, we perform a 3-fold Cross-Validation trying
different hyper-parameters configurations and 
obtaining the best performances on the validation set (i.e., average accuracy of $82.9\%$) using GloVe word embeddings and $128$ LSTM units\footnote[4]{For further details on the model selection and evaluation steps, the reader should refer to our previous works \cite{morini2020,morini2021toward}}. 
Consequently, we apply the model to the {\fontfamily{qcr}\selectfont r/politics} dataset in order to infer posts leaning for all the population. 
Notice that we apply it separately for each month in order to assess users' ideology stability over time.
For each textual content, we obtain model predictions ranging from $0$ to $1$ (i.e., the model confidence), where $1$ means that the post aligns itself with Republican (specifically Pro-Trump) ideologies while $0$ with Democrats (specifically Anti-Trump) ones. 
Then, for each user $u$ who participated in the debate in each month $m$ we compute his \textit{leaning score}, $L_{u, m}$, as the average value of his monthly posts leaning as follows:
\begin{equation} 
L_{u, m} = \frac{\sum_{p \in P_{u, m}} \text{Prediction\_Score}(p)}{\vert P_{u, m} \vert}
\label{form:polarization_score}
\end{equation}
where $P_{u, m}$ is the set of posts shared by a user $u$ in each month $m$. 
Since we are interested in assessing if users with polarized opinions tend to move to more moderate positions, we discretized such leanings into three intervals: \textit{Democrat} if $L_{u,m} \leq$ 0.4; \textit{Republican} if $L_{u,m} \geq$ 0.6; while \textit{Neutral} if $0.4 <  L_{u,m} < 0.6$. 
By adding a third label, we make sure to capture users with highly polarized ideologies.
Figure \ref{fig:user_leaning}b shows the authors' leaning score distribution for the whole time period obtained by averaging users' monthly scores. 
Such a distribution confirms what observed in a recent work \cite{de2021no} that focus the analysis on the {\fontfamily{qcr}\selectfont r/politics} subreddit too: Republican users (530,909) outnumber Democrat ones (185,256) and Neutrals users (373,630) show a tendency towards republicans beliefs.
\\ \ \\
\noindent{\bf Network Definition.}
Given the users for which we inferred their ideology, we define their interaction networks for each month to take into account the evolution of leanings in time. 
The resulting networks have Reddit users as node sets, $V$, and as edges the set of pair $(u,v) \in V$ for which a reply of $u$ to a $v$'s post or a comment exists.
We set each edge weight to represent the total number of comments exchanged between two users.  
Also, we label users (i.e., nodes) with their discretized \textit{leaning score} $L_{u,m}$ (i.e., Republican, Neutral and Democrat). 
In Table \ref{tab:network_statistics}, we provide the main averaged network statistics across the 20 considered months.
\begin{table}
\centering
\begin{tabular}{ccccccc} 
\hline
\textbf{$N$} & \textbf{$N_{R}$ }    & \textbf{$N_{D}$}     & \textbf{$N_{N}$}     & \textbf{ $E$ }       & \textbf{ $\left \langle k \right \rangle$ }         & \textbf{$r$ }         \\ 
\hline
183296       & 88285                & 36385                & 58626                & 1271080              & 6.97                 & 0.020                 \\ 
\hline
             & \multicolumn{1}{l}{} & \multicolumn{1}{l}{} & \multicolumn{1}{l}{} & \multicolumn{1}{l}{} & \multicolumn{1}{l}{} & \multicolumn{1}{l}{} 
\end{tabular}
\caption{Network statistics averaged across the 20 considered months: number of users $N$, divided in Republican $N_{R}$, Democrat  $N_{D}$ and Neutral $N_{N}$, number of edges $E$, network average degree $\left \langle k \right \rangle$, and network assortativity $r$ with respect to the political leaning.}
\label{tab:network_statistics}
\end{table}

\begin{figure}[t!]
\includegraphics[scale=0.17]{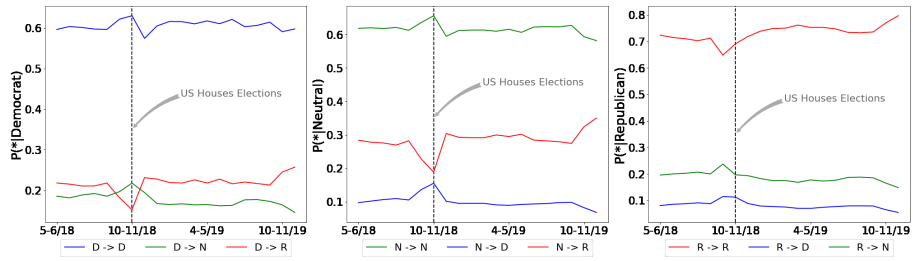}
\caption{For each ideology, users' transition probabilities over contiguous months.}
    \label{fig:trends}
\end{figure}

\section{Political Leaning: Stability and Open-Mindedness}\label{sec:stab_and_od}

\noindent{\bf Ideology Stability over Time.}
As a preliminary analysis, we try to understand how users' ideologies evolve. 
In detail, we are interested in exploring if users are stable and consistent with the same ideology or instead tend to change opinions according to specific events.
We model such an issue in terms of transition probabilities for such a purpose. 
In other words, for each user, we compute his probability $p_{ij}$ to move from \emph{state} $i$ to \emph{state} $j$ over contiguous months. 
Notice that, in this scenario, \emph{state} stands for the user's ideology (i.e., Republican, Neutral or Democrat). \newline
In Figure \ref{fig:trends} we show, for each ideology, the probability of users to change or remain in their states (i.e., opinions) over contiguous months.
At a first glance, we can observe a similar behavior for all the political ideologies: users tend to be rooted in their position over months, with a probability to remain in their state ranging around 0.60 for Democrat and Neutral leanings and around 0.70 for Republican. 
Moreover, Republican users - when changing their state - foster a more neutral position instead of moving to Democrat beliefs (differently from Democrats). 
These attitudes reflect the extreme polarization that characterizes the US political debate in the Trump Era \cite{bartels2018partisanship}.
However, for all the considered ideologies, we also notice an evident fluctuation of opinions in favor of Democrats and Neutrals around November 2018, precisely when (i.e., 6 November) the Democratic Party won control of the US House from the Republican Party.
Indeed, this was a decisive moment in which Democrats won the seats needed to take the House after capturing districts where President Trump was unpopular. 
\\ \
\begin{algorithm}[t]
\scriptsize
\begin{algorithmic}[3]
        \caption{\textbf{Confidence bound estimation algorithm.}\\
        $G_{t} = $ Weighted undirected interaction network at time t;\\ 
        $V_{t} = $ set of nodes at time t;\\
        $E_{t} = $ set of weighted edges at time t;\\
        $x_{u}(t)$ = opinion of agent $u$ at time $t$;\\
        $d_{u,v}$ = $|x_{u}(t)-x_{v}(t)|$ = opinion distance between $u,v\in V$ at time t;\\
        $\widehat{CB}$ = estimated confidence bounds.
        }
        \label{alg:cbest2}
            \If{$u \in V_{t+1}$}
                \State \text{Procedure to estimate} $\widehat{{x}_{u}}(t+1)$ and $\widehat{CB}_{u}$
                \State $N_{u,t} = \{ v | (u, v) \in E_{t}\}$; $|N_{u,t}| = n$
                \State $X_{N_{u, t}}[1...n] = \text{Array of opinions of nodes } v \in N_{u,t}$

                \If{ $N_{u,t} \neq \emptyset $ }
                    \State $\text{\text{Sort }} X_{N_{u, t}}[1...n] \text{ \textit{by} } d_{u,v} \text { \text{in ascending order.}}$
                    \State $\widehat{{X}_{u}}(t+1)[0...n] \text{ array of estimated opinions}$
                    \State $\widehat{{X}_{u}}(t+1)[0] = x_{u}(t)$
                    \State $E=[0...n] \text{ array of estimation errors}$
                    \State $E[0]=1.0$
                    \For{i=1; i=n; i++}
                        \State $x_{v} = X_{N_{u, t}}[i]$
                        \State $\widehat{{X}_{u}}(t+1)[i]= \frac{\widehat{X}_{u}(t+1)[i-1]+x_{v}}{2}$
                        \State $E[i]=|\widehat{{X}_{u}}(t+1)[i]-x_{u}(t+1)|$
                        \State $min_e=E[n]$
                        \For{i=n; i=0; i--}
                            \State $e=E[i]$
                            \If{$e \leq min_e$}
                                \State $min_e=e$
                                \State $j=i$
                            \EndIf
                        \EndFor
                    \EndFor
                    \State $\widehat{x_{u}}(t+1)=\hat{X}_{u}(t+1)[j]$
                    \State $\widehat{CB}=|x_{u}(t)-X_{N_{v, t}}[j]|$
                \EndIf
            \EndIf
        
\end{algorithmic}
\end{algorithm}    

\noindent{\bf Estimating Open-mindedness.}  
In order to understand the levels of open-mindedness involved in the process of changing one's political leaning, we started by assuming a simple process of opinion evolution at the individual level, based on a very well known model of opinion formation \cite{Deffuant2000MixingBA}. 
In opinion dynamics models, agents update their opinions after interacting with their neighbors according to simple mathematical rules. 
For example, in \cite{Deffuant2000MixingBA} agents average their opinion with the opinion of their interacting peer, which is randomly chosen from the pool of their neighbors, if and only if their opinion distance is below a certain threshold representing the open-mindedness of the population. 
The hypothesis that open-mindedness is a characteristic trait of an entire population and not a characteristic that varies from individual to individual is strong and probably unrealistic.
For this reason, in the present work, we assume a Deffuant-like process of opinion update (i.e., users averaging their opinions with the opinions of their interacting partners in a pairwise fashion) and provide a data-driven time-aware estimate of individual-level open-mindedness. 
To estimate users' tendency to be influenced by their neighbors, we developed a simple approach (see Alg. \ref{alg:cbest2} for implementation details) that takes as input the weighted interaction network at time $t$ and the opinions of the agents at time $t$ and time $t+1$. 
In the estimation procedure, we select each node $u$ for which we have both opinions $x_u(t)$ and $x_u(t+1)$ (Alg. \ref{alg:cbest2} line 1) so that we can estimate how much the interactions that happened in that time step influenced the opinion change and therefore obtaining an estimate for the level of bounded confidence. 
After selecting $u$, we order all the interacting partners (the neighbours of the node in the snapshot network) from the closer to the further by the opinion distance absolute value $d_{u,v}(t) = |x_u(t) - x_i(t)|$ (Alg. \ref{alg:cbest2} line 6). 
Then we compute an estimate $\widehat{x_u}(t+1)$ by iteratively averaging the new estimate with the interacting neighbours (Alg. \ref{alg:cbest2} line 13). 
The final estimated value $\widehat{x_u}(t+1)$ is the one that minimizes the error with respect to the real value $x_u(t+1)$ (Alg. \ref{alg:cbest2} lines 15-22). 
Finally, we compute the confidence bound as the distance in absolute value with the neighbor that represents the point of minimum in the estimation errors sequence (Alg. \ref{alg:cbest2} line 25). 
With the proposed approach, we can compute an estimate only for the subset of users present in two consecutive observations (i.e., months) and have at least one link (i.e., interaction) in the snapshot graph.

\begin{figure}[t]
    \centering
    \includegraphics[width=1\textwidth]{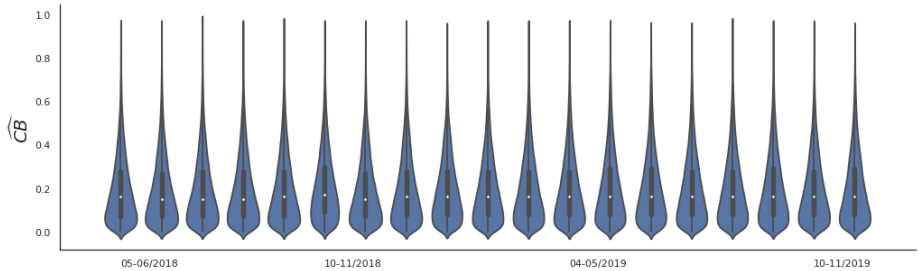}
    \caption{{Estimated open-mindedness $\widehat{CB}$ distribution from May-2018 to December-2019.} Distributions of the estimated confidence bound $\widehat{CB}$ over the whole time period. All distributions are positively skewed and constant over time.}
    \label{fig:cbest}
\end{figure}

\begin{figure}[t]
    \subfloat[]{\includegraphics[scale=0.32]{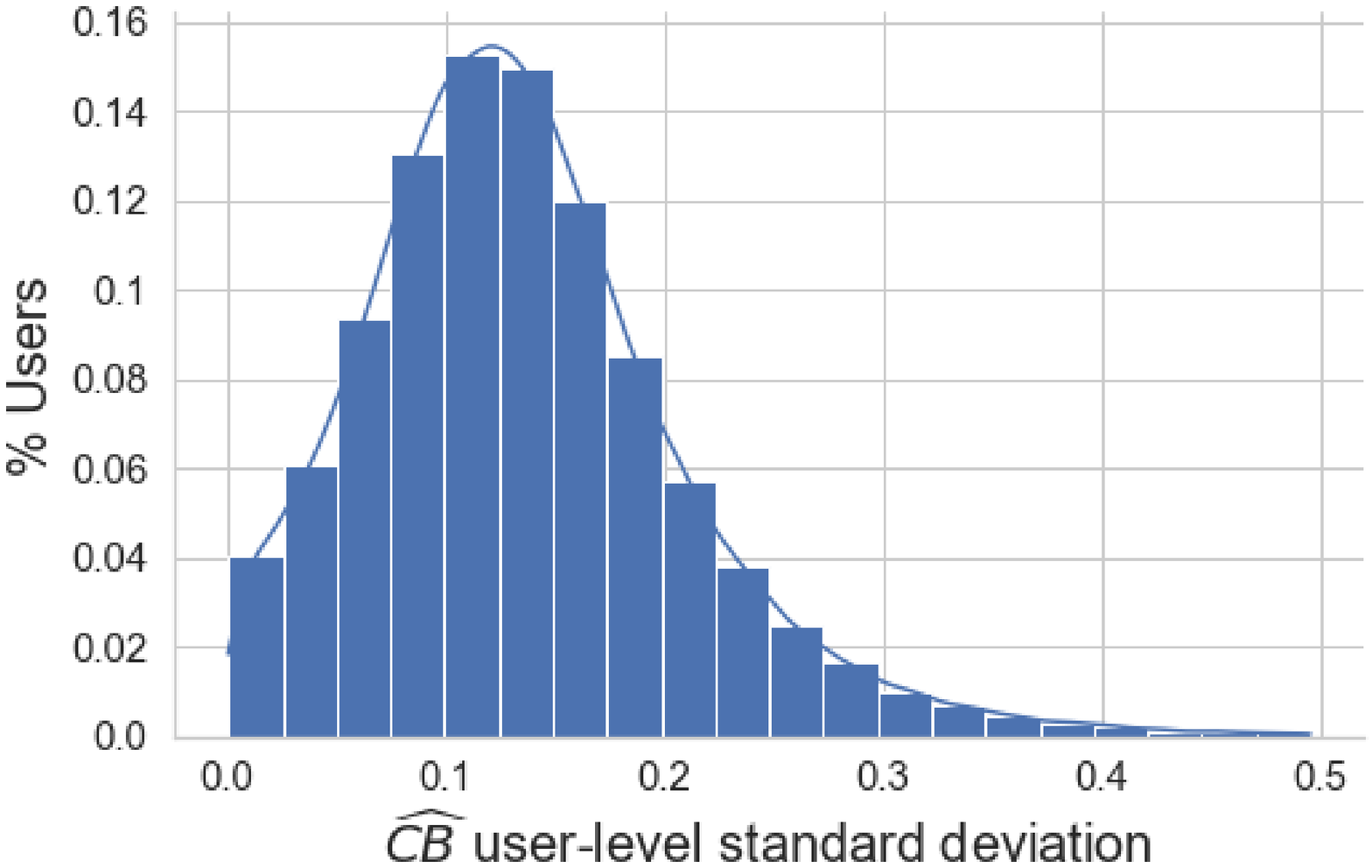}}
     \subfloat[]{\includegraphics[scale=0.32]{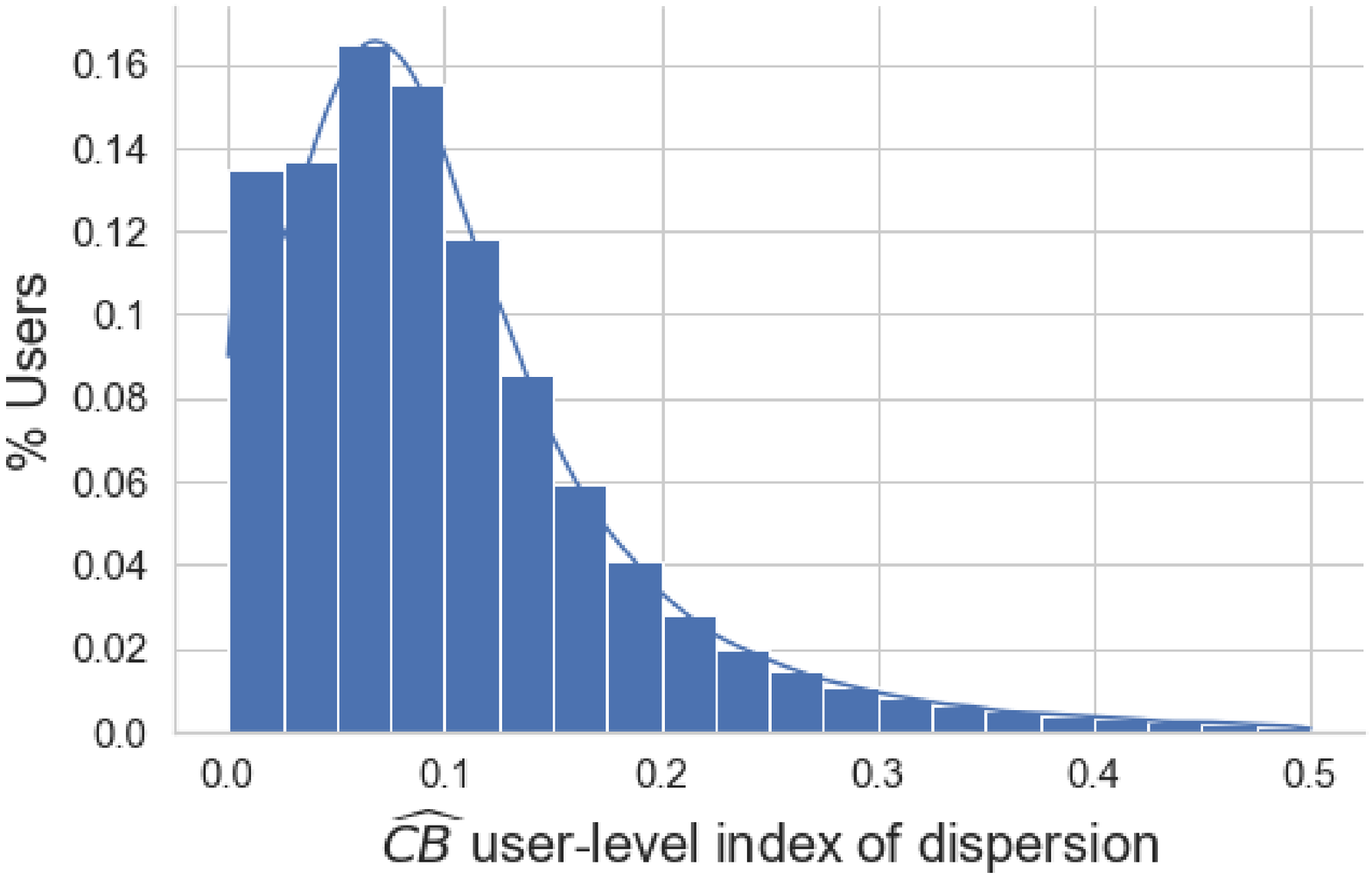}}
    \caption{Users' open-mindedness stability analysis. (a) Distribution of individuals' open-mindedness standard deviations; (b) Distribution of individuals' open-mindedness dispersion indexes (variance over mean value).}
    \label{fig:individual}
\end{figure}

\noindent Figure \ref{fig:cbest} underlines that the distribution of the estimated open-mindedness $\widehat{CB}$ is stable during the considered time period in the analyzed discussion. 
Moreover, it also highlights that the majority of the analyzed users is \say{close-minded}, i.e., their confidence bound is $\widehat{CB} \leq 0.2$, which is considered to be a sufficient condition for the population to become polarized in the long term, according to \cite{Deffuant2000MixingBA}. 
This means that most of the users participating in this discussion can hardly be influenced by neighbors holding distant opinions, even if they interact with these users during the considered time period, like in the case of this discussion, where the network has a low assortativity with respect to the political leaning (see network assortativity in Table \ref{tab:network_statistics}). 
However, we can also see that the distributions at each time step have a very high variance, allowing the presence of individuals having a level of open-mindedness close to $1.0$ - indicating that some of the users can also be influenced by people holding very different opinions and changing their expressed political leaning accordingly. 
The distribution of the estimated confidence bound for the opinions is less skewed between October and November 2018, i.e., around the US House of Representatives elections. 
Such behavior confirms the data-driven analysis based on transition probabilities performed in Section \ref{sec:data}: in this time window, Republicans, which normally have a highly skewed distribution, seemed to be somehow more open-minded, and their average confidence bound is higher.
\begin{figure}[t]
\includegraphics[width=\textwidth]{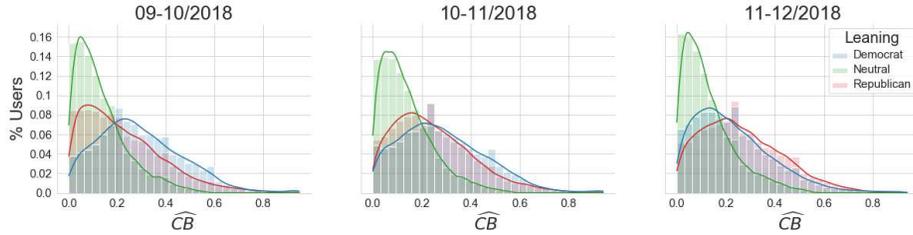}
\caption{Estimated open-mindedness distribution in the period September - December 2018. Distributions of the estimated confidence bound $\widehat{CB}$ for the different political leanings: Democrat (blue), Neutral (green), and Republican (red) from September to December 2018.}
\label{fig:3months_cb}
\end{figure}
In figure \ref{fig:3months_cb} we reported the estimated distribution for Democrat, Neutral, and Republican users (we took the orientation at time t and estimated the confidence bound $\widehat{CB}$ between time t and t+1). 
Only three months are present in this figure, i.e., the three months around the US House election, but conclusions still hold for the other months considered in this work.
We can see from figure \ref{fig:3months_cb} that there are differences in the distribution of open-mindedness when we consider political orientation. 
Both neutral and republican distributions are positively skewed, and the distribution has a long right tail. 
Also, the Democrats distribution has a right tail, but much less than the others. 
From our analysis, it appears that Neutral individuals are also the most close-minded, while Democrats have a wider range of confidence bound levels. 
Their distribution is not as skewed as the others, and many users have a very high level of bounded confidence, i.e., they change their opinion significantly over contiguous time periods under the influence of their neighbors. 
Republicans, like Neutrals, have a positively skewed confidence bound distribution, even with a higher average confidence bound. 
We performed a 2-sample KS-test comparing the distributions of the estimated confidence bound for each political leaning (e.g., Democrat vs. Republican $\widehat{CB}$) within each time step obtaining a p-value $\approx 0.0$, supporting the conclusions that distributions are different for the three political leanings. 
Finally, while we can say that population-level open-mindedness is reasonably constant over time, i.e., we have the same mix of open-minded and close-minded individuals participating in the discussion, we do not have information about how variable open-mindedness may be at the individual level. 
In this analysis, each user may have a different value of open-mindedness at each time step, making not only the overall distribution heterogeneous but also the distribution at the individual-level. 
To understand how much individual bounded confidence may vary, we computed the standard deviation of our estimate for each observed user - figure \ref{fig:individual}a.
Reddit users' open-mindedness tendencies appear stable in time, showing a characteristic low standard deviation, $\sigma$.
Such a result is also confirmed by the distribution of the Fano dispersion index (figure \ref{fig:individual}b) - i.e., the ratio between variance $\sigma^2$ and mean value, $\mu$, of the estimated individual open-mindedness scores.
The observed Fano values, prevalently distributed in the range $0 < \frac{\sigma^2}{\mu} < 1$, identify under-dispersed behaviors, thus expressing consistent patterns of stability.

\section{Conclusions}
\label{sec:conclusions}
In this work, we introduced a methodology to estimate individuals' open-mindedness from online discussion data.
When applied to study the political debate that took place on Reddit during the first two years of the Trump presidency, our proxy unveiled the existence of characteristic distributions for well-defined sub-populations: Moderates/Neutrals and Republicans being more close-minded on average than Democrats.
The proposed longitudinal analysis also unveiled that the observed Reddit users tend to maintain a stable behavior for what concerns their open-mindedness, exhibiting low variance and underdispersion.

Indeed the current study, like all empirical ones, suffers from limitations. 
In particular, it leverages a data-driven estimate of the political leaning that can be subject to errors and cannot be fully validated on ground truth external data.
Moreover, although Reddit users tend to be particularly involved in political discussions, the population variability in time and the sparsity of observation data do not allow to estimate the open-mindedness of less active individuals. For each month, we were not able to estimate the open-mindedness of $40-50\%$ of nodes since we have no information on the opinion at time $t+1$ or the user does not have neighbors on the network at time $t$. 

As future works, we plan to enhance the proposed estimation procedure, allowing for asymmetric open-mindedness.
Additionally, we will investigate the interplay of open-mindedness and known polarization phenomena (e.g., the presence of echo chambers) in order to better characterize the role of different individuals in their emergence.

\subsubsection*{Acknowledgments}
This work was supported by the scheme 'INFRAIA-01-2018-2019: Research and Innovation action', Grant Agreement n. 871042 'SoBigData++: European Integrated Infrastructure for Social Mining and Big Data Analytics, by the CHIST-ERA grant CHIST-ERA-19-XAI-010, by MUR (grant No. not yet available), FWF (grant No. I 5205), EPSRC (grant No. EP/V055712/1), NCN (grant No. 2020/02/Y/ST6/00064), ETAg (grant No. SLTAT21096), BNSF (grant No. o 
\begin{otherlanguage*}{russian}КП-06-ДОО2/5\end{otherlanguage*}
)



\end{document}